\documentclass[11pt,a4paper,svgnames]{article}
\pdfoutput=1

\usepackage{latexsym}
\usepackage{epsf}
\usepackage{verbatim}
\usepackage{xfrac}
\usepackage{slashed}
\usepackage{tikz}
\usepackage{cancel}
\usepackage{bbm}

\usepackage{fancyhdr}
\usepackage{datetime}

\usepackage{makeidx}

\renewcommand{\Re}{\textrm{Re}}
\renewcommand{\d}{\textrm{d}}

\newcommand{\Tr}{\textrm{Tr}}

\newcommand{\w}{\wedge}

\newcommand\varpm{\mathbin{\vcenter{\hbox{%
  \oalign{\hfil$\scriptstyle+$\hfil\cr
          \noalign{\kern-.3ex}
          $\scriptscriptstyle({-})$\cr}%
}}}}

\newcommand\varmp{\mathbin{\vcenter{\hbox{%
   \oalign{\hfil$\scriptstyle-$\hfil\cr
           \noalign{\kern-.3ex}
          $\scriptscriptstyle({+})$\cr}%
}}}}

\usetikzlibrary{arrows,automata,positioning,calc,trees,decorations.pathmorphing,decorations.markings}

\usepackage[utf8]{inputenc}
\usepackage{uniinput}       % Type unicode directly, including in math mode!

\usepackage{amsmath, amsfonts, amssymb, amsthm}
\usepackage{mathtools}      % For \coloneqq
\usepackage{graphicx}
\usepackage{hyperref}
\usepackage{url}
\usepackage{booktabs, multirow, bigdelim}       % For nice looking tables
\usepackage{authblk}        % For author affiliations
\usepackage{csquotes}       % To use \enquote{…} instead of ``…''.
\usepackage[left=1in,right=1in,top=1in,bottom=1.25in]{geometry}
\usepackage{microtype}      % For pretty typography, configure options as needed.
\usepackage{xcolor}
\usepackage[plain]{fancyref}
\usepackage{perpage}
\usepackage{soul}
\usepackage{siunitx, dcolumn}
\usepackage{colortbl}
\usepackage{cite, mcite}
\usepackage[title]{appendix}
\usepackage{tocloft}

\usepackage[format=plain,
textfont=it]{caption}
\hypersetup{colorlinks, breaklinks,
    urlcolor = DarkBlue,
    linkcolor = DarkBlue,
    citecolor = DarkBlue}

\renewcommand{\Im}{\rm{Im}}

\colorlet{soulyellow}{yellow!30}
\colorlet{soulgreen}{green!30}

\newcolumntype{d}[1]{D{.}{.}{#1}}

\newcommand{\ie}{\textit{i.e.~}}
\newcommand{\eg}{\textit{e.g.~}}

\newcommand{\be}{\begin{equation}}
\newcommand{\ee}{\end{equation}}
\newcommand{\ben}{\begin{displaymath}}
\newcommand{\een}{\end{displaymath}}
\newcommand{\bea}{\begin{eqnarray}}
\newcommand{\eea}{\end{eqnarray}}

\newcommand{\bean}{\begin{eqnarray*}}
\newcommand{\eean}{\end{eqnarray*}}

\newcommand{\inlinechi}[2]{\raisebox{\depth}{$#1\chi$}} % inner command, used by \rchi
\DeclareRobustCommand{\rchi}{{\mathpalette\inlinechi\relax}}

\MakePerPage{footnote}

\usepackage{alphalph}
\makeatletter
\newalphalph{\fnsymbolwrap}[wrap]{\@fnsymbol}{}
\makeatother

\def\@fnsymbol#1{\ensuremath{\ifcase#1\or \dagger\or \ddagger\or
        \mathsection\or \mathparagraph\or \|\or **\or \dagger\dagger
        \or \ddagger\ddagger \else\@ctrerr\fi}}

\graphicspath{{figures/}}

\renewcommand{\d}{\textrm{d}}

\title{\bfseries Constructing stable de Sitter in M-theory from higher curvature corrections}
\author[1]{Johan Blåbäck\footnote{johan.blaback@roma2.infn.it}}
\author[2]{Ulf Danielsson\footnote{ulf.danielsson@physics.uu.se}}
\author[2]{Giuseppe Dibitetto\footnote{giuseppe.dibitetto@physics.uu.se}}
\author[2]{Suvendu Giri\footnote{suvendu.giri@physics.uu.se}}

\affil[1]{Dipartimento di Fisica \& Sezione INFN, Università di Roma ``Tor Vergata'',
	Via della Ricerca Scientifica 1, 00133 Roma, Italy}
\affil[2]{Institutionen för fysik och astronomi,
	Uppsala Universitet, Box 803, SE-751 08 Uppsala, Sweden}

\date{} % Don't print a date

\begin{document}
    % !TEX root = main.tex

\begin{flushright}
    \texttt{ROM2F/2019/02}\\
    \texttt{UUITP-6/19}
\end{flushright}
\vskip -20pt

\begingroup
    \let\newpage\relax      % Void the actions of \newpage
    \maketitle
\endgroup

    \vskip 30pt
    % !TEX root = main.tex

\begin{abstract}
    \noindent
    %------------------------------------------------------%
    {\normalsize We consider dimensional reductions of M-theory on $\mathbb{T}^{7}/\mathbb{Z}_{2}^{3}$ with the inclusion of arbitrary metric flux and spacetime filling
    KK monopoles. With these ingredients at hand, we are able to construct a novel family of non-supersymmetric yet tachyon free Minkowski extrema. 
    These solutions are supported by pure geometry with no extra need for gauge fluxes and possess a fully stable perturbative mass spectrum, up to
    a single flat direction. Such a direction corresponds to the overall internal volume, with respect to which the scalar potential exhibits a no-scale
    behavior.
    We then provide a mechanism that lifts the flat direction to give it a positive squared mass while turning $\textrm{Mkw}_{4}$ into
    $\textrm{dS}_{4}$. The construction makes use of the combined effect of $G_{7}$ flux and higher curvature corrections. Our solution is scale separated and the quantum corrections are small.
    Finally we speculate on novel possibilities when it comes to scale hierarchies within a given construction of this type, and possible issues with the choice of quantum vacuum.}
    %------------------------------------------------------%
\end{abstract}
    \newpage
    \tableofcontents
	% !TEX root = main.tex

\section{Introduction}
\label{sec:introduction}
Ever since the turn of the millennium, we have known that our universe is currently undergoing a phase of accelerated expansion driven by the so-called dark energy.
The compelling experimental evidence for such a cosmic acceleration is independently corroborated by measurements involving type Ia supernovae \cite{Riess:1998cb,Perlmutter:1998np}, the Cosmic Microwave Background (CMB) radiation \cite{Jaffe:2000tx} and the Baryonic Acoustic Oscillations (BAO) \cite{Tegmark:2003ud}.
A combination of these observational data resulted in the $\Lambda$CDM model of cosmology, which describes the dark energy content of our universe by means of a positive and small cosmological constant.

From the perspective of high energy physicists and string theorists in particular, this has posed the challenge of embedding cosmic acceleration
within a UV complete description of gravity such as string theory. The avenue which has been mostly pursued so far is that of finding metastable de Sitter (dS) solutions
directly within string theory. The mechanism of flux compactification seemed at the time a very promising tool to achieve this.
However, despite some initial enthusiasm, it turned out to be remarkably difficult to provide rigorous constructions possessing dS vacua within a weakly coupled regime.

In fact, all the constructions which are currently available in the literature involve ingredients that are arguably not well under control.
Among these, the most successful mechanism is the one proposed by KKLT \cite{Kachru:2003aw}, where the use of non-perturbative effects is a key ingredient.
Their starting point  is a ``no-scale'' Minkowski vacuum obtained within type IIB string theory \cite{Giddings:2001yu}, in which the universal K\"ahler modulus encoding information concerning the internal volume is completely unfixed.
Subsequently, by invoking non-perturbative effects, one can argue for a new contribution to the potential that non-trivially depends on this field.
The final step then makes use of antibranes to lift the cosmological constant to a positive value, thus constructing the desired metastable dS extremum.
However, as already anticipated earlier, employing exotic ingredients such as non-perturbative effects or antibranes, has been extensively argued to give rise to
instabilities of various sorts. See \cite{Danielsson:2018ztv} for a review of the difficulties one may encounter when attempting similar constructions.

On the other hand, stringy de Sitter constructions in general have recently been subject to some skepticism at a much more fundamental level.
Possibly inspired by the unitarity and entropy puzzles raised by dS geometry when written in its static patch, one may start wondering whether a healthy theory
of quantum gravity should even allow for dS vacua in the first place. In this context, the authors of \cite{Obied:2018sgi} came up with a ``swampland dS conjecture'',
according to which static dS extrema should be ruled out in any string compactifications.
While the conjecture in its original form of \cite{Obied:2018sgi} provides an $\mathcal{O}(1)$ bound on the first slow-roll parameter of any scalar potential
arising from a string compactification, the more recent refined versions only put constraints on either the first or the second slow-roll parameter
 \cite{Ooguri:2018wrx} (in line with the original proposals of \cite{Andriot:2018wzk,Garg:2018reu,Garg:2018zdg}). Note that this possibility still leaves room
 for realizing cosmic acceleration in the form of quintessence, although the bounds are fairly narrow \cite{Akrami:2018ylq, Raveri:2018ddi}.

Swampland arguments aside, finding an explicit and convincing example of a dS vacuum in string theory, realistic or not, still represents a longstanding challenge for string cosmologists.
Going back to the no-scale Minkowski vacuum, as an alternative to non-perturbative effects, one may consider adding non-geometric fluxes to get some non-trivial
dependence on the K\"ahler moduli. Such models do indeed possess metastable dS extrema \cite{Font:2008vd,deCarlos:2009qm,Danielsson:2012by,Blaback:2013ht,Damian:2013dwa}.
However, even though such generalized fluxes were first introduced in \cite{Shelton:2005cf} based on string duality arguments, a clear perturbative understanding of these
objects is still lacking.

Yet another somewhat appealing alternative is that of considering higher derivative corrections to the Einstein-Hilbert term.
Thanks to non-renormalization theorems, these corrections are known to only contribute to a shift in the K\"ahler potential, while the flux induced superpotential remains blind to these effects. 
This gives rise to a new dependence on the internal volume that could in principle fix it.
This strategy has been extensively studied in the context of type IIB compactifications (see \eg \cite{Berg:2004ek,Berg:2005yu}), where the first non-trivial higher derivative
corrections come from an $\alpha'^{3}$ term containing eight derivatives\footnote{Lower orders in $\alpha'$ have been argued not to contribute, at least at tree and 1-loop level in the string coupling.}.
This was computed both by supergravity \cite{Green:1998by,Rajaraman:2005up,Paulos:2008tn} and matched by string worldsheet techniques \cite{Berg:2005ja,Berg:2007wt} for some orbifolds relevant to
flux compactifications.

The aim of this paper is to consider reductions of M-theory on a 7-manifold with the inclusion of higher derivative contributions.
The form of such corrections was studied in \cite{Antoniadis:1997eg,Green:1997di,Green:1997as,Russo:1997mk}, where the next-to-leading term is found to be given by an $\mathcal{R}^{(4)}$ term, together with higher derivative terms involving $G_4$ factors  related to it by supersymmetry and contributing at the same order. 
Our starting point will be a novel class of Minkowski solutions that are obtained when reducing M-theory on a flat 7D group manifold, with the extra addition
of spacetime filling KK monopoles. These solutions possess a non-negative perturbative mass spectrum with the only flat direction being the volume modulus.
We are then in the position of invoking higher curvature terms to lift this direction. The advantage of choosing these Minkowski vacua
as a starting point, rather than the no-scale solutions in IIB, is that we have a \emph{purely real} dilatonic flat direction, while in the usual no-scale
models the flat direction is a complete \emph{complex} scalar modulus. This implies that, even after lifting the volume by using higher derivative terms, one would then still be left with
an axionic flat direction that can only be lifted by further combining these with non-perturbative effects.
Contrary to the IIB setup, we will not need them here.

What we find are fully stable, scale separated dS vacua at large volume. A curious and possibly essential feature, which we elaborate on in the last section, is that the supersymmetry breaking scale set by the gravitino mass, $m_{3/2}$, is {\it higher} than the Kaluza-Klein cutoff scale. This means that the spontaneously broke supersymmetry of the low energy $\mathcal{N}=1$ supergravity theory is not restored before the extra dimensions become important. In particular, the superpartners of standard model particles are not part of the low energy theory.  In the last section, we also observe that the quantum corrections are small. Nevertheless, we do point out possible issues related to the choice of quantum vacuum. It will be interesting to confront our proposed dS vacuum with the various versions of the swampland conjectures. 

Remaining within the M-theory setup, which is relevant to our present work, it is interesting to put our results into the context of \cite{Acharya:2006ia,Acharya:2008zi,Kane:2011kj}, where it is claimed that phenomenologically appealing particle physics models can be obtained from M-theory compactifications on $\textrm{G}_2$ manifolds. 
In particular, in \cite{Acharya:2007rc}, and more recently in \cite{Kane:2019nod} with applications to inflation, it is argued that their analysis can be combined with the achievement of a stable dS vacuum through the use of non-perturbative terms without the need of extra uplifting ingredients such as \eg antibranes. Similar constructions directly yielding dS have been discussed in \cite{Blaback:2013qza,Danielsson:2013rza,Guarino:2015gos,Blaback:2015zra} using non-geometric as well as non-perturbative terms. In particular \cite{Blaback:2015zra} discusses how non-perturbative terms can be mimicked by non-geometric terms and vice versa. 
In this sense though, the special feature of the KKLT construction is that the non-perturbative terms produce a supersymmetric AdS vacuum, which then is uplifted to a dS.
The proposal of this paper is radically different than all of the above examples, since it does not invoke any exotic ingredients. It would be extremely interesting to further explore our construction and its compatibility with realistic particle physics.

The paper is organized as follows. In \Fref{sec:minkowski}, we present a new class of non-supersymmetric Minkowski solutions arising from M-theory compactified on a G\textsubscript{2} structure manifold. The main feature of these solutions is that they have the overall volume as a flat direction. In \Fref{sec:curvature}, we first discuss the general form of higher derivative corrections to 11 dimensional supergravity and later specify to the case of our interest. We then propose in \Fref{sec:ds} a mechanism to lift the flat direction using the above corrections, the result being a metastable de Sitter extremum. Finally, in \Fref{sec:loop} we present some general discussion and conclusive remarks, mainly focused on the hierarchies of the different physical scales involved.

	% !TEX root = main.tex

\section{The tree-level Minkowski vacuum}
\label{sec:minkowski}

M-theory compactifications on twisted toroidal orbifolds have been widely studied in the last few decades \cite{Dasgupta:1995zm,Acharya:1998pm}. In the special case of $\mathbb{T}^{7}/\mathbb{Z}_{2}^{3}$ \cite{DallAgata:2003txk,DallAgata:2005zlf},
the effective four dimensional description is given by minimal supergravity coupled to seven chiral multiplets, which specify the data of the
internal manifold's $\textrm{G}_{2}$ structure. In a weakly coupled regime, these models turn out to describe type IIA reductions on $\mathbb{T}^{6}/\mathbb{Z}_{2}^{2}$,
with additional spacetime filling O6 planes. In this limit, the scalar fields within the seven chiral multiplets are to be interpreted as the string
coupling, the complex structure, and the K\"ahler moduli, respectively.

At a general level, models of this type realize $\textrm{SL}(2,\mathbb{R})^{7}$ global bosonic symmetry. The scalar sector consists of seven complex fields $\Phi^{\alpha}\,\equiv\,\left(S,T_{i},U_{i}\right)$ with $i=1,2,3$.
In the present work we shall focus on the so-called \emph{isotropic} sector of the theory obtained by identifying the three $T_{i}$ as well as the
three $U_{i}$ moduli.
In this case, the kinetic Lagrangian for the remaining three (universal) complex fields follows from the K\"ahler potential
\be
\label{eq:Kaehler_STU}
\mathcal{K}\,=\,-\log\left[-i\,(S-\overline{S})\right]\,-\,3\log\left[-i\,(T-\overline{T})\right]\,-\,3\log\left[-i\,(U-\overline{U})\right]\ ,
\ee
$\mathcal{L}_{\textrm{kin}}=\mathcal{K}_{α\bar{β}}\,∂Φ^α ∂\bar{Φ}^{\bar{β}}$, where $\mathcal{K}_{α\bar{β}} \coloneqq \partial_{\alpha}\partial_{\bar{\beta}}\mathcal{K}$ is the Kähler metric.
The inclusion of fluxes and a non-trivial metric twist induces a scalar potential $V$ for the would be moduli fields, which may be written as
\be
\label{eq:V_N=1}
V\,=\,e^{\mathcal{K}}\left(-3\,|\mathcal{W}|^{2}\,+\,\mathcal{K}^{\alpha\bar{\beta}}\,D_{\alpha}\mathcal{W}\,D_{\bar{\beta}}\overline{\mathcal{W}}\right)\ ,
\ee
in terms of the above K\"ahler potential and an arbitrary holomorphic superpotential $\mathcal{W}$, where $\mathcal{K}^{\alpha\bar{\beta}}$ represents the inverse K\"ahler metric and $D_{\alpha}$ denotes the K\"ahler-covariant derivative.

Though the explicit form of the superpotential is not further constrained by minimal local supersymmetry, the one induced by M-theory fluxes within this particular orbifold is given by \cite{DallAgata:2005zlf}
\be
\label{eq:W_Geom}
\mathcal{W}\,=\,a_0-b_0 S+3 c_0 T-3 a_1 U+3 a_2 U^2+3 b_1 S U+3(2 c_1 -\tilde{c}_1)T U +3 c_3^{\prime} T^2-3 d_0 S T\ .
\ee
The eleven dimensional origin of the above superpotential terms as M-theory fluxes is given in \Fref{tab:fluxes}, while in \Fref{tab:sources}
the associated flux tadpoles are discussed, in connection with the corresponding spacetime filling sources.

\begin{table}
    \centering
            \begin{tabular*}{0.8\textwidth}{@{\extracolsep{\fill}}cccc}
                \toprule
                \sc{couplings} & \sc{m-theory} & \sc{type iia} & \sc{fluxes} \\ \midrule
                $1 $ &  $G_{ambncp7}$ & $F_{ambncp}$ & $  a_0 $ \\
                $S $& $ {G}_{mnp7} $ & $ {H}_{mnp} $ & $  -b_0$ \\
                $T $& $ G_{a b p7} $ & $ H_{a b p} $ & $  c_0 $ \\
                $U $ &  $G_{ambn}$ & $F_{ambn}$ & $   -a_1 $ \\ \arrayrulecolor{gray}\midrule
                $U^{2} $ &  ${{\omega}_{am}}^{7}$ & $F_{am}$ & $  a_2 $ \\
                $S \, U $ &  ${{\omega}_{mn}}^{c}$ & ${{\omega}_{mn}}^{c}$ & $  b_1 $ \\
                $T\, U $ &  $ {\omega_{p a}}^{n} = {\omega_{b p}}^{m} \,\,\,,\,\,\, {\omega_{b c}}^a $  & $ {\omega_{p a}}^{n} = {\omega_{b p}}^{m} \,\,\,,\,\,\, {\omega_{b c}}^a $ & $c_1 $, $\tilde{c}_1$ \\
                $T^{2}$ &  ${{\omega}_{a 7}}^{m}$ & \rm{non-geometric} & $  c_3^{\prime} $ \\
                $S\, T $ &  $ {\omega_{7 m}}^{a} $  & \rm{non-geometric} & $-d_0 $ \\
                \arrayrulecolor{black}\bottomrule
            \end{tabular*}
\caption{The relation between M-theory fluxes and superpotential couplings. One should note that the distinction between the roles of the three
different moduli $S$, $T$ and $U$ only becomes relevant when taking the weak coupling limit to type IIA.
In this limit, the direction labeled by ``$7$'' is singled out, while the other six naturally split into $3+3$ (respectively labeled by ``$a$'' \& ``$m$''), which
are even (odd) with respect to the orientifold involution enforced by \emph{O6}$^{||}$ planes. The remaining \emph{O6}$^{\perp}$ planes realize the $\mathbb{Z}_{2}^{2}$ orbifold.
\label{tab:fluxes}}
\end{table}
\begin{table}[h!]
    \centering
            \begin{tabular*}{0.8\textwidth}{@{\extracolsep{\fill}}ccc}
                \toprule
                \sc{m-theory sources} & \sc{iia sources} & \sc{tadpoles}  \\ \midrule
                (KK6/KKO6) & (O6/D6)$^{||}$ &  $6a_2 b_1$ \\
                (KK6/KKO6) & (O6/D6)$^{\perp}$ &  $-a_2(2c_1-\tilde{c}_{1})$  \\ \arrayrulecolor{gray}\midrule
                (KK6/KKO6) & (KK5/KKO5) &  $a_2 d_0-b_1 (c_1-\tilde{c}_{1})$ \\
                (KK6/KKO6) & (KK5/KKO5) & $a_2 c_3^{\prime}-c_1(c_1-\tilde{c}_{1})$ \\
                \arrayrulecolor{gray}\midrule
                (KK6/KKO6) & \rm{exotic} & $-c_3^{\prime}(2c_1-\tilde{c}_{1})$ \\
                (KK6/KKO6) & \rm{exotic} & $b_1 c_3^{\prime}+d_0 (c_1-\tilde{c}_{1})$ \\
                (KK6/KKO6) & \rm{exotic} & $2 (b_1 c_3^{\prime}+2 c_1 d_0)$ \\
                \arrayrulecolor{black}\bottomrule
            \end{tabular*}
\caption{The flux tadpoles supporting the M-theory flux backgrounds in \Fref{tab:fluxes} and their weakly coupled type IIA
interpretation. Eleven dimensional KK monopoles turn out to produce either \emph{D6} branes or type IIA KK monopoles, when reduced on the isometry direction
or a worldvolume direction, respectively. However, they may give rise to exotic objects with no clear IIA description when reduced along a transverse direction.
\label{tab:sources}}
\end{table}

\subsection*{Finding tachyon free Minkowski extrema}

The starting point is to observe \cite{Dibitetto:2011gm} that the search for critical points of the scalar potential arising from \eqref{eq:W_Geom} can be restricted to the
origin of the scalar manifold, \ie  where $S=T=U=i$, without loss of generality. This comes at the price of keeping the flux parameters in \Fref{tab:fluxes}
completely arbitrary, since any non-compact $\textrm{SL}(2,\mathbb{R})^{3}$ transformation needed to move any point to the origin would just reparametrize
the superpotential couplings in \eqref{eq:W_Geom}. This translates the original problem into that of solving six algebraic quadratic equations in the flux parameters.

The next step needed to further simplify the search and identify our family of non-supersymmetric flat Minkowski vacua, is to make use of the Maldacena-Nu\~nez
no-go theorem \cite{Maldacena:2000mw} specified to our case. Practically speaking, one just rewrites the scalar potential (at the origin) as a linear combination of the field equations,
plus some residual part which has a definite sign
\begin{equation}
\label{eq:Maldacena_Nunez_Identity}
V|_{S=T=U=i}\,=\, \lambda^{I}\,\partial_{I}V|_{S=T=U=i} \, + \, V_{0} \ ,
\end{equation}
where $\lambda$ is a suitable vector of constants, the contracted index $I$ runs over the six real scalars\footnote{The explicit parametrization that we adopt throughout this work is
$S=\chi+i\,e^{-\varphi}$, $T=\chi_{1}+i\,e^{-\varphi_{1}}$ and $U=\chi_{2}+i\,e^{-\varphi_{2}}$, while $I\in\{\chi,\,\varphi,\,\chi_{1},\,\varphi_{1},\,\chi_{2},\,\varphi_{2}\}$.}, and $V_{0}$ is a suitable negative definite quadratic expression in the fluxes. This procedure yields
\begin{equation}\label{eq:lambdaV0}
\lambda\,=\,\left(0,\,\frac{1}{3},\,0,\,1,\,0,\,1\right)\ , \qquad V_{0}\,=\, -\frac{1}{48} \left(2 a_0^2+3 a_1^2+b_0^2+3 c_0^2\right) \ ,
\end{equation}
which implies that $V|_{\textrm{on-shell}}\leq\,0$, thus ruling out dS extrema.

Besides ruling out dS vacua, an identity of the form in \eqref{eq:Maldacena_Nunez_Identity} can be used as a hint to where in parameter space one should
look for Mkw solutions, \ie  in those regions where $V_0$ vanishes identically. In our specific case, taking \eqref{eq:lambdaV0} into account, one is easily
lead to the conclusion that Mkw solutions (if there are any!) are confined within a region of parameter space where no gauge fluxes are turned on, but just pure metric flux.
Therefore, we will look for critical points in the origin of moduli space, where furthermore
\begin{equation}
a_0 \ = \ 0\ , \  a_1 \ = \ 0 \ , \ b_0 \ = \ 0 \ , \ c_0 \ = \ 0 \ ,
\end{equation}
where we can also restrict to the choice $\tilde{c}_{1}=c_1$, since the both parameters contribute to the same superpotential term.
The only non-trivial constraints for Mkw extrema in this case are
\begin{equation}
V|_{\textrm{origin}}\ \overset{!}{=} \ 0\ , \ \partial_{\varphi}V|_{\textrm{origin}}\,=\,\partial_{\varphi_{1}}V|_{\textrm{origin}}\,=\,\partial_{\varphi_{2}}V|_{\textrm{origin}}\ \overset{!}{=} \ 0 \ .
\end{equation}
The above system of algebraic equations is homogenous, and hence one can always use an overall rescaling to fix \eg $d_0=1$.
After such a procedure, there turn out to exist four independent one dimensional branches of solutions, two of which are supersymmetric and identical, while the
other two are non-supersymmetric and still tachyon free for appropriate choices of points.
A numerical example thereof is obtained by performing the flux choice in \Fref{tab:minkowski_fluxes} and vanishing gauge fluxes.
\begin{table}[]
    \centering
    \begin{tabular*}{0.6\textwidth}{@{\extracolsep{\fill}}c*{1}{S[table-format = 1.15]}}
        \toprule
        \multicolumn{1}{c}{\sc{fluxes}} & \multicolumn{1}{c}{\sc{minkowski fluxes}} \\ \midrule
        $a₂$    &  -0.007083446561363 \\
        $b₁$    &  -0.153111518425194 \\
        $c₁,\tilde{c}₁$    & 0.140508634503070 \\
        $c₃^\prime$ & -0.908052565020441 \\
        $d₀$    & 0.999999997219754 \\ \bottomrule
    \end{tabular*}
    \caption{\label{tab:minkowski_fluxes}
        Choice of fluxes that generates a non-supersymmetric Minkowski critical point. All of the above flux parameters are metric flux components, while the gauge fluxes are zero.}
\end{table}

Remarkably, the above Mkw solution is completely free of tachyons, even when checking the non-isotropic directions. This can be explicitly seen in \Fref{tab:mkw_mass_spectrum},
where the complete non-isotropic mass spectrum is computed for our numerical example. As for flat directions, there turns out to be just one,
whose existence is directly implied by identity \eqref{eq:Maldacena_Nunez_Identity} whenever setting $V_0=0$. The actual direction is determined by the
vector $\lambda$, and in this particular case, it corresponds to the sum of all the dilatons of the theory.
\begin{table}[]
    \centering
    \begin{tabular*}{0.8\textwidth}{@{\extracolsep{\fill}}cSS}
        \toprule
        & \sc{\qquad isotropic} & \sc{\qquad non-isotropic} \\ \midrule
        \multirow{3}{*}{\sc{axionic directions:}} & 1.025171 & 0.122146 $\quad(\times 2)$ \\
        & 0.158575 & 0.000673 $\quad(\times 2)$ \\
        & 0.001076 &  \\ \arrayrulecolor{gray}\midrule
        \multirow{3}{*}{\sc{dilatonic directions:}} & 0.500241 & 0.454137 $\quad(\times 2)$ \\
        & 0.001362 & 0.010290 $\quad(\times 2)$ \\
        & 0 &  \\ \arrayrulecolor{black}\bottomrule
    \end{tabular*}
\caption{The complete (non-isotropic) physical mass spectrum for our flat Minkowski solution in \Fref{tab:minkowski_fluxes}. The numerical values are calculated as the eigenvalues of $\mathcal{K}^{JK} \, D_{I} \, D_{K}V|_{\mathrm{sol.}}$.
Note that the eigenvalues with double multiplicity correspond to the non-isotropic modes, while the only exactly flat direction represents the overall internal volume.
\label{tab:mkw_mass_spectrum}}
\end{table}

Physically, the class of Minkowski solutions obtained in this way may be viewed as a dimensional reduction of M-theory on a flat group manifold of dimension seven, with
the addition of spacetime filling KK monopoles. Alternatively, by virtue of \cite{Danielsson:2014ria}, the effect of including KK monopoles can be resolved into geometry by
promoting the internal manifold to a more general $\textrm{G}_{2}$ structured manifold.
In either case, the overall internal volume remains completely unfixed at this stage, hence we need to supplement this construction with a mechanism that
lifts the this flat direction by including new effects which depend on the volume.

	% !TEX root = main.tex

\section{Higher derivative corrections in M-theory}\label{sec:curvature}

The low energy limit of M-theory is described by eleven dimensional supergravity, whose only dimensionful gravitational coupling is given by the 11D Planck length
$\ell_{11}$. This theory effectively describes type IIA strings at finite string coupling. A dimensional reduction of M-theory on a circle of radius $R_{11}$
yields type IIA string theory, where the 10D string length $\alpha'\equiv\ell_{s}^{2}$ and the string coupling are determined by
\begin{equation}
\begin{array}{lcccclc}
\alpha' \ = \ \dfrac{\ell_{11}^{3}}{2\pi R_{11}} & , & & \textrm{and } & \ g_{s}^{2} \ = \ \dfrac{2\pi R_{11}^{3}}{\ell_{11}^{3}} & .
\end{array}
\label{IIA_M_couplings}
\end{equation}
Starting from a perturbative ($g_{s}\ll 1$) type IIA description, for which the cutoff scale for higher derivative corrections is set by $\Lambda_{10}\sim\left(2\pi\alpha'\right)^{-1/2}$,
one might think of going to finite string coupling up in eleven dimensions and determine the 11D cutoff scale from simple dimensional analysis through
\begin{equation}
\Lambda_{11}^{3}\,R_{11} \ \overset{!}{=} \ \Lambda_{10}^{2} \ \overset{\eqref{IIA_M_couplings}}{\longrightarrow} \ \Lambda_{11} \, = \, \ell_{11}^{-1} \ ,
\end{equation}
which is in particular completely independent of the compactification radius $R_{11}$. This implies that the 11D Planck length may be used to construct
a whole tower of local and diffeomorphism invariant higher derivative corrections to 11D supergravity. This would give rise to $L$-loop contributions to the effective action of the form \cite{Russo:1997mk}
\begin{equation}
\label{Tower_Loop_11dcorrections}
\delta S_{L} \, = \, \kappa_{11}^{2(L-1)}\,\sum\limits_{(n,\ell,m)}\Lambda_{11}^{n}\left(\log \Lambda_{11}\right)^{\ell}\,\int d^{11}x \sqrt{-g_{11}}\,\mathcal{R}_{11}^{(m)} \ ,
\end{equation}
where $2\kappa_{11}^{2}\equiv(2\pi)^{5}\ell_{11}^{9}$, $\mathcal{R}_{11}^{(m)}$ denote some suitable curvature invariants involving $2m$ derivatives, and the integer
labels $(n,\ell,m)$ must satisfy the following constraints
\begin{equation}
\label{nmL_constraint}
n \, + \, 2m \, \overset{!}{=} \, 9(L-1) \, + \, 11 \ ,
\end{equation}
by virtue of dimensional analysis.
By specializing \eqref{Tower_Loop_11dcorrections} to one loop and furthermore truncating the log expansion to zero-th order, we find the following 1-loop effective action
\begin{equation}
\label{Tower_1Loop_11dcorrections}
S_{1\textrm{-loop}}\,=\,(2\pi)^{-5}\int d^{11}x \sqrt{-g_{11}}\,\sum\limits_{m=1}^{\infty}\ell_{11}^{2m-11}\mathcal{R}_{11}^{(m)} \ ,
\end{equation}
where $\mathcal{R}_{11}^{(1)}\equiv\mathcal{R}_{11}$, \ie  simply the 11D Ricci scalar. However, the above tower of 11D higher curvature corrections is constrained through its relation to type IIA stringy corrections. In particular, if we start from a $2m$ derivative term in type IIA, this will generically be of the form
$f(g_{s})\mathcal{R}_{10}^{(m)}$, where $f$ represents an arbitrary function of the string coupling, which we assume to asymptotically behave as $g_{s}^{2(k-1)}$ as
$g_{s}\gg 1$. By performing a circle reduction of \eqref{Tower_Loop_11dcorrections}, one finds
\begin{equation}
\kappa_{11}^{2(L-1)}\Lambda_{11}^{n}\,\sqrt{-g_{11}}\,\mathcal{R}_{11}^{(m)} \ \overset{\eqref{nmL_constraint}}{\longrightarrow} \ g_{s}^{\frac{2}{3}(m-4)}\,\mathcal{R}_{10}^{(m)} \ ,
\end{equation}
which \emph{only} matches the stringy higher genus expansion $g_{s}^{2(k-1)}$ for integer $k$'s if $m\,=\,3k+1$.

In summary, the first non-trivial 1-loop higher derivative corrections to 11D supergravity are argued to occur at $m=4$, \ie
\begin{equation}
\label{Leading_1Loop_11dcorrections}
S_{1\textrm{-loop,NLO}}\,=\,\frac{1}{2\kappa_{11}^{2}}\,\int d^{11}x \sqrt{-g_{11}}\,\left(\mathcal{R}_{11}\,+\,\kappa_{11}^{4/3}\,\mathcal{R}_{11}^{(4)}\right) \ ,
\end{equation}
where the explicit form of the eight-derivative curvature invariant $\mathcal{R}_{11}^{(4)}$ is given by \Fref{eq:explicit_R4}.
For a thorough discussion of the form of the full eight-derivative action of M-theory, including higher order terms in $G_{4}$, we refer to Appendix~\ref{app:appendix}.

Before moving to the background specific form of the corrections when reducing on a $\textrm{G}_{2}$ structure manifold, it is perhaps worthwhile making a last general comment.
We have seen how a term of the form $\kappa_{11}^{2(L-1)}\Lambda_{11}^{n}\mathcal{R}_{11}^{(m)}$ may produce a type IIA $L$-loop correction if $m=3L+1$.
Besides this, $\mathcal{R}_{11}^{(3L+1)}$ will also contribute at tree level in type IIA through a ``Casimir-type'' term obtained by replacing \cite{Russo:1997mk}
\begin{equation}
\Lambda_{11}^{n} \ \longrightarrow \ \left(\Lambda_{11}^{n}\,+\,c\,R_{11}^{-n}\right) \ \sim \ \left(\mathcal{O}(g_{s}^{2L-2})\,+\,c\,\mathcal{O}(g_{s}^{-2})\right) \ ,
\end{equation}
which is indeed a sum of an $L$-loop and a tree level term in $g_{s}$.

\subsection*{Corrections to M-theory on G\textsubscript{2} structure manifolds}

Let us now specify to the concrete class of M-theory backgrounds preserving $\mathcal{N}=1$ supersymmetry down in four dimensions, which are known as $\textrm{G}_{2}$ structures. Our logic will be that of first understanding the case
of 7-manifolds with $\textrm{G}_{2}$ holonomy, \ie torsion free $\textrm{G}_{2}$ structures, and their relation to CY 3-fold backgrounds in type IIA.
Only later we will move on to the situation of $\textrm{G}_{2}$ structures with non-trivial torsion, which is relevant to our work.

Thanks to non-renormalization theorems, higher derivative corrections to compactifications preserving 4D minimal supersymmetry can only contribute to renormalizing
the effective K\"ahler potential of the theory, while the superpotential is protected by holomorphicity arguments from receiving perturbative corrections of any sort.
On general grounds, corrections to the lower dimensional effective action are expected to only depend on the actual field content of the full quantum theory.
For CY compactifications in type IIA, this turns out to be encoded in the topological data of the internal manifold, \ie the Hodge numbers $h^{(1,1)}$ \& $h^{(2,1)}$ that specify
the amount of vector and hypermultiplets, respectively.
By taking an orientifold of the these models, we end up with an effective $\mathcal{N}=1$ supergravity where both the vectors and the hypers reduce to chiral multiplets.
The full set of $\alpha'$ corrections can then be encoded in the so-called \emph{quantum-corrected volume}
replacing the classical volume modulus in the K\"ahler potential. This effect produces a $(\textrm{vol}_{6})^{-1}$ correction to the K\"ahler potential at leading order,
whose coefficient is proportional to the Euler character $\rchi(\mathcal{M}_{6})$ of the compact 6-manifold.
In \cite{Grana:2014vva} both $\alpha'$ and $g_{s}$ corrections were studied for type IIA reductions on manifolds with $\textrm{SU}(3)$ structure.
Remarkably, the exact form of the corrections was argued to be completely independent of the $\textrm{SU}(3)$ torsion and thus identical to the torsion free $\textrm{CY}_{3}$ case.

Moving now to $\textrm{G}_{2}$ structures, we again borrow our intuition from \cite{Grana:2014vva} and expect to learn something about higher derivatives
from the torsion free case. This means that the problem of determining the higher derivative corrections to the minimal theory arising from M-theory on $\mathbb{T}^{7}/\mathbb{Z}_{2}^{3}$ dressed up
with $\textrm{G}_{2}$ torsion, is related to understanding the resolution of such an orbifold to obtain a smooth 7-manifold with $\textrm{G}_{2}$ holonomy \cite{Shatashvili:1994zw}.
Once this is explicitly constructed, the quantum-corrected volume appearing in the K\"ahler potential will be governed by the following topological quantity
\begin{equation}
\xi(\mathcal{M}_{7}) \, \equiv \, 7b^{0} \, - \, 5b^{1} \, + \, 3b^{2} \, - \, b^{3} \ ,
\end{equation}
where the $\left\{b^{i}\right\}$'s are the Betti numbers of the 7-manifold.
Note that such an invariant also happens to reduce to the usual Euler character when evaluated for special 7-manifolds obtained as $\mathcal{M}_{6}\times S^{1}$:
\begin{equation}
\xi(\mathcal{M}_{6}\times S^{1})  \, = \, \rchi(\mathcal{M}_{6}) \ .
\end{equation}
The above object was first introduced in \cite{Duff:2010ss} as an odd dimensional
generalization of the Euler character that has the property of flipping sign under generalized mirror symmetry. Moreover, the topological invariant
$\xi$ also appears to determine the total on-shell 4D trace anomaly through
\begin{equation}
g_{\mu\nu}\,\langle T^{\mu\nu}\rangle \, = \, \frac{1}{32\pi^{2}}\,\left(-\frac{\xi}{24}\right)\,E_{4} \ ,
\end{equation}
where $E_{4}$ denotes the Euler density.

Returning to the construction of smooth resolutions of the $\mathbb{Z}_{2}^{3}$ toroidal orbifold, this can be found in the original constructive proof
of \cite{Joyce:1996_1,Joyce:1996_2}. The approach there consists in deforming the original singular orbifold with $\textrm{G}_{2}$ holonomy by a ``smoothing off'' parameter
denoted by $t$. This smoothing procedure generically turns on a torsion. However, it is then shown that taking the $t\rightarrow 0$ limit still defines a new
perfectly smooth compact manifold with $\textrm{G}_{2}$ holonomy. This argument as such further corroborates our intuition that the Betti numbers of
our actual $\textrm{G}_{2}$ structure manifold will stay the same as the ones computed in the case of vanishing torsion.

In the $\mathbb{Z}_{2}^{3}$ toroidal orbifold at hand, the original singular manifold has $\left(b^{0},\,b^{1},\,b^{2},\,b^{3}\right) \, = \, \left(1,\,0,\,0,\,7\right)$,
thus yielding a vanishing $\xi$. The $\mathbb{Z}_{2}^{3}$ orbifold action turns out to have eight singular points, the structure of the singularities being
$\mathbb{T}^{3}\times\left(\mathbb{C}^{2}/\mathbb{Z}_{2}\right)$. The smoothing procedure then consists in replacing each of the singularities by patches of
$\mathbb{T}^{3}\times U_{j}$, where $U_{j}$ is a regular 4-manifold with $\textrm{SU}(2)$ holonomy that asymptotes to $\mathbb{C}^{2}/\mathbb{Z}_{2}$. These patches might then need to
be quotiented by some discrete involution $F_{j}$, acting non-trivially on the 3-torus as well.

In the case of interest to us, $U_{j}$ is the Eguchi-Hanson space \cite{Eguchi:1978xp}, while the needed involution is a $\mathbb{Z}_{2}$.
However, its action on the patch $\mathbb{T}^{3}\times U_{j}$ can be chosen in \emph{two topologically inequivalent} ways. This arbitrary choice in each of our
eight orbifold singularities results in $2^{8}=256$ in principle different 7-manifolds, of which in the end only nine are inequivalent.
The inequivalent choices can be labeled by the integer $p=0,\,\dots,8$ counting how many singularities were smoothed out by picking the first choice for the
involution. The corresponding Betti numbers of the manifold obtained by this prescription ($\mathcal{M}^{(p)}_{7}$), and the relation to the matter content of the resulting 4D quantum description are collected in\footnote{In the table ``GM'' $=$ ``Gravity Multiplet'',
``VM'' $=$ ``Vector Multiplet'', and ``$\rchi$M'' $=$ ``Chiral Multiplet''.}  \Fref{tab:Betti}.
\begin{table}
    \centering
    \begin{tabular*}{0.8\textwidth}{@{\extracolsep{\fill}}ccccc}
        \toprule
        \sc{betti numbers}: & $b^{0}$ & $b^{1}$ & $b^{2}$ & $b^{3}$ \\
        \midrule
        \sc{matter multiplets}: & GM & not allowed & VM & $\rchi$M \\
        $b^{i}(\mathcal{M}^{(p)}_{7})$: & $1$ & $0$ & $(8+p)$ & $(47-p)$ \\ \bottomrule
    \end{tabular*}
    \caption{\label{tab:Betti}
        The explicit relation between the topological data of a smooth $\textrm{G}_{2}$ manifold and the matter content of the effective $4$D description.
        The actual values of the Betti numbers in the last line refer to the inequivalent smoothing procedures labeled by the integer $p$.}
\end{table}
The resulting $\xi$ for the aforementioned resolved versions of $\mathbb{T}^{7}/\mathbb{Z}_{2}^{3}$ reads
\begin{equation}
\xi(\mathcal{M}^{(p)}_{7}) \, = \, 4(p-4) \ , \qquad \textrm{with } \ p=0,\,\dots,8\ ,
\end{equation}
which, in particular can be of \emph{either signs}, depending on the value of $p$.

The final step is now to write down the general form of the corrections to the K\"ahler potential, and we are particularly interested in their scaling behavior
with respect to the volume modulus, since our final goal is that of stabilizing it by adding such higher curvature effects to the flat Minkowski extremum that we found previously.
This is done by evaluating the 1-loop effective action \eqref{Leading_1Loop_11dcorrections} on a background with natural $4+7$ splitting
\begin{equation}\label{eq:metric}
ds_{11}^{2} \, = \, \rho^{-7/2}\, ds_{4}^{2} \, + \, \rho\, ds_{\mathcal{M}_7}^{2} \ ,
\end{equation}
where $\rho$ is the volume modulus, while $\mathcal{M}_7$ is taken to be unit volume.
Note that the $\rho^{-7/2}$ factor in front of the 4D metric is needed to yield the 4D Einstein frame upon reduction.
At this point, part of the $\mathcal{R}_{11}^{(4)}$ terms will be of the form $\mathcal{R}_{4}\,\mathcal{R}_{7}^{(3)}$, thus giving rise to
\begin{equation}
\sqrt{-g_{4}}\,\mathcal{R}_{4}\,\underbrace{\left(1\,+\, c \, \xi(\mathcal{M}_7)\,\kappa_{11}^{4/3} \, \rho^{-3}\right)}_{\equiv \, M_{\textrm{Pl}}^{2}} \ ,
\end{equation}
where $c$ is a numerical constant which will not be crucial to our analysis. The above correction to the effective 4D Planck mass may in turn be
viewed as a correction to the K\"ahler potential $\mathcal{K}$, which reads
\begin{equation}
\label{eq:K1loop}
\mathcal{K}_{\textrm{1-loop}} \, = \, \mathcal{K}_{\textrm{tree}} \ - \ c \, \xi(\mathcal{M}_7)\,\kappa_{11}^{4/3} \, \rho^{-3} \ ,
\end{equation}
where it should be stressed that the only thing that will be crucial in the analysis we perform in the next section, is the fact
that $\xi(\mathcal{M}_7)$ can be chosen to have the right sign, and that the above expression still holds true once a non-trivial $\textrm{G}_{2}$ torsion is turned on.

	% !TEX root = main.tex

\section{Constructing stable dS}\label{sec:ds}

The one-loop corrected Kähler potential of \Fref{eq:K1loop} can be written as
\begin{equation}
    \mathcal{K}_\textrm{1-loop} = \mathcal{K}_\textrm{tree} + Δ\mathcal{K} = \mathcal{K}_\textrm{tree} - \frac{e₁}{ρ³}\,,
\end{equation}
where $e₁ \coloneqq c \xi(\mathcal{M}_7) \kappa_{11}^{4/3}$.
%%%%%%%%%%%%%%%
The addition of $\Delta \mathcal{K}$ generates a leading contribution (in a large $\rho$ expansion) to the scalar potential of the form $ΔV \sim - e₁ρ^{-15/2}$. Further corrections to the scalar potential of the form $\sim ρ^{-21/2}$ can be generated by switching on a 7-form field strength $G_7$. These can be used together to stabilize the flat direction of \Fref{sec:minkowski}.
We will discuss this in detail below.

The $\mathcal{R}^{(4)}$ corrections and $G_7^2$ contribution are expected to shift the Minkowski solution form the values in \Fref{sec:minkowski}. In anticipation of this shift, we can perturb the fluxes so that the new superpotential reads
\begin{equation}
    \begin{split}
        \mathcal{W}   &= \mathcal{W}_0 + Δ\mathcal{W} \\
            &= \mathcal{W}_0 + \left(a₀ + δa₂ U² + δb₁ S U + δc₁ T U + δc₃^\prime  T² + δd₀ S T \right)\\
            &= a₀ + \left(3a₂ + δa₂\right) U² + \left(3b₁ + δb₁\right) S U + \left(3c₁+δc₁\right) T U +\\ &\quad\left(3c₃^\prime + δc₃^\prime \right) T² + \left(-3d₀ + δd₀\right) S T.
    \end{split}
\end{equation}
The $a_0$ term here corresponds to addition of the $G_7$ flux, while the other fluxes has been perturbed as $f \to f + \delta f$, where their 11 dimensional interpretation is listed in \Fref{tab:fluxes}.

We choose the fluxes ($a₀, b₁, c₁, c₃^\prime$ and $d₀$, which are collectively denoted by $f$), to be of the same order (say $N$), while $a₀$ is chosen to be of order $N²$ and $ρ$ is taken to be of order $N^{2/3}$ \ie
\begin{equation}\label{eq:Nscaling}
a_0 = a_{00} N^2\,,\quad \rho^3 = \rho_0^3 N^2\,,\quad f = f_0 N.
\end{equation}
Taking $N$ to be large corresponds to taking the large volume limit (\ie large $ρ$), while keeping the fluxes (\ie $f$) large so that the perturbations $δf$ are automatically small ($1/N$ suppressed) compared to $f$.
The contribution of these terms to the scalar potential, and their effective scaling with respect to $N$ (up to factors of the fluxes determined by the Minkowski solution that we perturb around), are summarized it \Fref{tab:potential_contributions}.

In the expansion parameter $N$, the Minkowski solution is established to order $1/N$, meaning that the potential in the absence of $e_1$, $a_0$, and $\delta f$ is zero up to an overall factor of $N^{-1} ρ₀^{-9/2}$. For large $N$, the second order term in $e_1$ \ie $e₁²N²ρ^{-21/2}$, is therefore subleading compared to all other terms. The term $δb₁Nρ^{-9/2}$ is generated from expressions of the schematic form $\mathcal{W}Δ\mathcal{W}$, and can be made to vanish by choosing a particular relation between the flux perturbations. In the absence of this term, the potential goes as $N^{-3}$ up to leading order and is a sum of positive terms ($\sim a₀²$ and $\sim δb₁²$) and a negative term ($\sim - e₁$).
The parameters can then be chosen such that the potential has a dS minimum.
Let us do this in detail.

To summarize, the perturbed potential in the large $N$ expansion has the form
\begin{equation}
  V = \frac{1}{N} \widetilde{V}_{\textrm{Mkw}} + \frac{1}{N^2} \widetilde{V}_{\delta f} + \frac{1}{N^3} (\widetilde{V}_{a_{00}^2} + \widetilde{V}_{(\delta f)^2} + \widetilde{V}_{e_1}) + \frac{1}{N^4} (\widetilde{V}_{e_1 a_{00}} + \widetilde{V}_{e_1 \delta f}) + \mathcal{O}(N^{-5})\,,
\end{equation}
where a tilde means we have extracted all the $N$-dependence from those terms, and $\widetilde{V}_x$ means (except for $\widetilde{V}_{\textrm{Mkw}}$) that the term has $x$ as a coefficient.
\begin{table}[]
    \centering
    \begin{tabular*}{0.8\textwidth}{c @{\extracolsep{\fill}}ccc}
        \toprule
        $Δ\mathcal{K}$  & $Δ\mathcal{W}$    & $ΔV$ & $N$ \sc{scaling} \\ \midrule
        \multirow{2}{*}{$e₁ρ^{-3}$} & \multirow{2}{*}{0} & $\hspace{-10pt}- e₁N²ρ^{-15/2}$ & $\hspace{-8pt}-N^{-3}$ \\
        &                    & $e₁²N²ρ^{-21/2}$ & $N^{-5}$ \\ \arrayrulecolor{gray}\midrule
        0&            $a₀$        & $a₀²ρ^{-21/2}$  &  $N^{-3}$ \\ \arrayrulecolor{gray}\midrule
        \multirow{2}{*}{0} & \multirow{2}{*}{$δb₁ S U$} & $δb₁²ρ^{-9/2}$ & $N^{-3}$ \\
        &                    & $δb₁Nρ^{-9/2}$ & $N^{-2}$ \\ \arrayrulecolor{black}\bottomrule
    \end{tabular*}
\caption{Contribution of the various terms to the scalar potential. Only one of the flux perturbations ($δb₁$) has been listed here for ease of representation. The effect of turning on perturbation to the other fluxes in identical.\label{tab:potential_contributions}}
\end{table}

We then proceed to derive our solution, which would be approximately de Sitter in the $N \gg 1$ range, by considering expressions order-by-order in this expansion. To leading order ,\ie $1/N$, we already have a solution which is the Minkowski solution. To second order, \ie $1/N^2$, we solve each equation of motion which is a set of three linear equations for our six flux perturbations, which originates from terms of the form $\mathcal{W}\Delta \mathcal{W}$ that give rise to the $\widetilde{V}_{\delta f}$ terms. The potential is still zero to $1/N^2$ order after this step.
To third order, \ie $1/N^3$, we choose the remaining parameters such that the potential is positive at its minimum, and solves the equation of motion along the $\rho$ direction to this order.

\begin{figure}[t]
    \centering
    \includegraphics{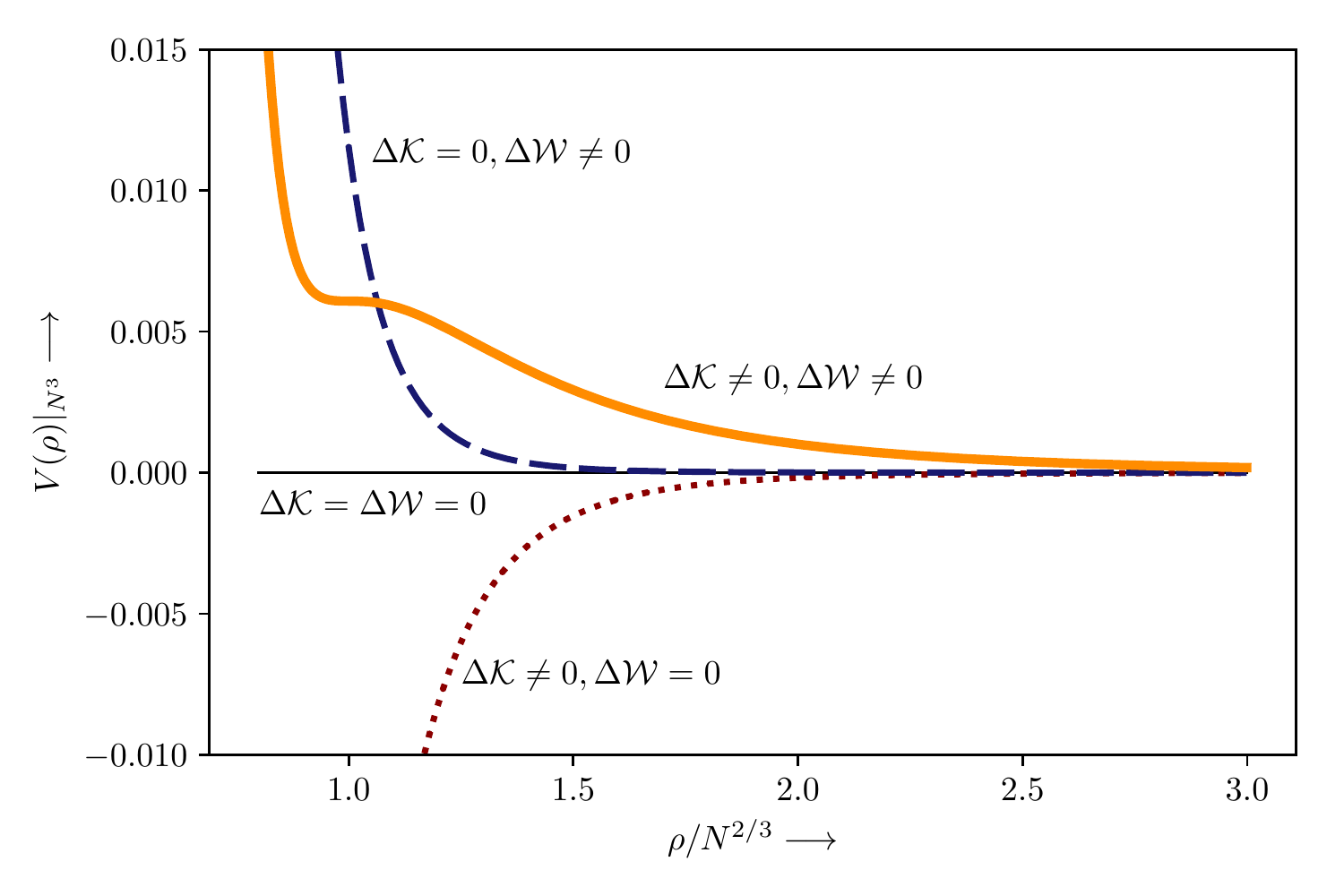}
    \caption{Potential $($truncated to $O(N^{-3}))$ along the volume modulus $ρ$. The potential generated by the higher curvature term in the Kähler potential is shown with the red dotted line and that of the additional terms in the superpotential is shown with the dashed blue line.
        The parameters in $Δ\mathcal{K}$ and $Δ\mathcal{W}$ can be chosen so that the total potential, represented by the thick orange line, has a metastable dS minimum.
        \label{fig:rhoplot}}
\end{figure}

Taking $e₁ = 1$ for example, perturbation to the fluxes can be chosen as in \Fref{tab:flux_perturbations} to stabilize the $ρ$ direction with a positive vacuum energy
\begin{equation}\label{eq:dSpotential}
  V = + \frac{0.006}{N^3} + \mathcal{O}(N^{-4})\,;\quad \partial_{ρ₀} V = \mathcal{O}(N^{-4})\,;\quad \partial_{ρ₀}^2 V = \frac{0.0006}{N³} + \mathcal{O}(N^{-4})
\end{equation}
at $\rho_0 = 1$.
\begin{table}[]
    \centering
    \begin{tabular*}{0.8\textwidth}{@{\extracolsep{\fill}}c*{2}{S[table-format = 1.17]}}
        \toprule
        \multicolumn{1}{c}{\sc{fluxes}} & \multicolumn{1}{c}{\sc{minkowski fluxes}} & \multicolumn{1}{c}{\sc{perturbations}} \\ \midrule
        $a_{00}$   & 0 & 0.604970163885781 $N²$ \\
        $a₂$    &  -0.007083446561363 $N$  & 0.034000000614343           \\
        $b₁$    &  -0.153111518425194 $N$ & -0.049999998095880       \\
        $c₁$    & 0.140508634503070 $N$  & 0.066186914364069          \\
        $c₃^\prime$ & -0.908052565020441 $N$  & -4.667599421364908           \\
        $d₀$    & 0.999999997219754 $N$  & 4.926293439693896           \\ \bottomrule
    \end{tabular*}
    \caption{\label{tab:flux_perturbations}
    	Choice of fluxes corresponding to a stable de Sitter minimum. Fluxes for the Minkowski vacuum from \Fref{tab:minkowski_fluxes} are scaled with $N$ and perturbed with order $1$ numbers. In addition, a constant term $a₀$ , proportional to $N²$ is added to the superpotential.}
\end{table}
By our choice of flux parameters at the $1/N^3$ level, the solution has been perturbed slightly in the directions transverse to $\rho$. Since these directions were already massive before our perturbation, they are still massive to leading order, but the exact position of the minimum has shifted. The new minimum can be determined to the correct order by opening up the other two dilatonic transverse directions $ψ₁,ψ₂$ by
choosing\footnote{Note that this choice of parametrization is precisely such that the corrected K\"ahler potential purely depends on the volume modulus $\rho$. Generically the explicit form of such corrections might carry a dependence on the transverse dilatonic directions, as well. However, such a dependence will not spoil our perturbative argument here, but would rather only slightly correct the actual position of the dS extremum by $N$ suppressed contributions.}
\begin{equation}\label{eq:st3u3}
    S = i(ρ ψ₁^{-3} ψ₂^{-3})^{3/2}\ , \quad T = i(ρ ψ₁)^{3/2}\ , \quad U = i(ρ ψ₂)^{3/2}\ .
\end{equation}
The $\psi_a$ have a perturbation from $1$ starting at $1/N²$ and their solution for our example is
\begin{equation}
  \psi_1 = 1 + \frac{13.931}{N²} + \mathcal{O}\left(N^{-3}\right)\ ,\quad \psi_2 = 1 -\frac{18.064}{N²} + \mathcal{O}\left(N^{-3}\right)\ .
\end{equation}
Finally we have an approximate de Sitter solution for which the scalar potential goes as $V\sim +\mathcal{O}\left(N^{-3}\right)$, while the equations of motion scale as $∂_iV\sim \mathcal{O}\left(N^{-4}\right)$, which gives the first slow-roll parameter as dependence of $N$ as
\begin{equation}
  \epsilon \coloneqq \frac{\mathcal{K}^{IJ}∂_IV∂_JV}{2V²} \sim \frac{2168.6}{N^2} + \mathcal{O}\left(N^{-3}\right)\ .
\end{equation}
The second slow-roll is given by the $\rho$ direction and is defined as
\begin{equation}
    η \coloneqq \textrm{Min.~Eig.}\left(\frac{\mathcal{K}^{IK} ∂_K∂_J V}{\left|V\right|}\right)\, \sim 0.01005 - \frac{134.78}{N} + \mathcal{O}(N^{-3})\ .
\end{equation}
It is worth noting that in this perturbative analysis in large $N$, we have assumed that $ρ₀=1$. However, in reality, when the potential along the heavy moduli $ψ_a$ shifts slightly so that the new minimum is not exactly at $ψ_a=1$ but at $ψ_a = 1+\mathcal{O}(N^{-2})$, $ρ₀$ is also expected to shift slightly from 1. A more exact analysis taking into account this backreaction can be done by doing a numerical computation parallel to the perturbative analysis above. Starting with the fluxes in \Fref{tab:flux_perturbations}, the equations of motion for the dilatons (\ie their combinations along the heavy directions $ψ_a$ and light direction $ρ$) can be solved which yields
\begin{equation}
    ρ₀ = \SI{0.993647398}, \quad ψ₁ = \SI{1.000000035}, \quad ψ₂ = \SI{0.999999953}\ .
\end{equation}
Since the equations of motion are solved exactly, it is an exact de Sitter minimum. The first slow roll parameter $\epsilon = 0$, while the second slow roll parameter $η \sim 0.010$. The full mass spectrum, normalized with respect to the volume \ie $\mathcal{K}^{IK} ∂_K∂_J V/\left|V\right|$, for $N=20 000$ is shown in \Fref{tab:dS_mass_spectrum}. Comparing it to the Minkowski mass spectrum in \Fref{tab:mkw_mass_spectrum}, we see that the heavy directions are the same as before up to scaling (note that the fluxes are scaled with $N$ which scales the Minkowski masses by $N²$. Additionally, unlike the Minkowski case, the masses here are normalized with respect to the potential). The potential along the volume modulus $\rho$ is plotted in figure \ref{fig:rhoplot}.
\begin{table}[]
    \centering
    \begin{tabular*}{\textwidth}{@{\extracolsep{\fill}}ccccccc}
        \toprule
        \sc{mass $\rightarrow$}& \SI{6.74e10} & \SI{1.04e10} & \SI{7.07e7} & \SI{3.29e10} & \SI{8.96e7} & $1.005\times 10^{-2}$ \\ \midrule
        \rotatebox[origin=c]{90}{\sc{eigenvector}}\hskip 2pt$\begin{pmatrix} χ\\φ\\χ₁\\φ₁\\χ₂\\φ₂ \end{pmatrix} \rightarrow$ &
        $\begin{pmatrix} 0.723\\0\\0.690\\0\\0.005\\0 \end{pmatrix}$ &
        $\begin{pmatrix} -0.931\\0\\0.324\\0\\0.165\\0 \end{pmatrix}$ &
        $\begin{pmatrix} 0.355\\0\\-0.131\\0\\0.925\\0 \end{pmatrix}$ &
        $\begin{pmatrix} 0\\0.953\\0\\-0.303\\0\\-0.015 \end{pmatrix}$ &
        $\begin{pmatrix} 0\\-0.479\\0\\-0.536\\0\\0.695 \end{pmatrix}$ &
        $\begin{pmatrix} 0\\0.577\\0\\0.577\\0\\0.577 \end{pmatrix}$  \\
        & \multicolumn{3}{c}{$\xleftrightarrow{\hskip 180pt}$} & \multicolumn{3}{c}{$\xleftrightarrow{\hskip 180pt}$} \\
        & \multicolumn{3}{c}{\sc{axionic directions}} & \multicolumn{3}{c}{\sc{dilatonic directions}} \\
        \bottomrule
    \end{tabular*}
    \caption{Mass spectrum of the de Sitter solution for the numerical example discussed in \Fref{sec:ds}. The complex fields S, T and U are parametrized as in \eqref{eq:st3u3}, $e₁=1$ and $N=2\times10⁴$.
        \label{tab:dS_mass_spectrum}
    }
\end{table}

To summarize, we are able to establish de Sitter solutions starting from the Minkowski solutions found in Section~\ref{sec:minkowski}. This is possible thanks to the particular property of this Minkowski solution where it has one flat direction left over, which we can stabilize using a perturbative correction and a $G_7$ contribution. This particular Minkowski solution is in turn possible thanks to the inclusion of KK monopoles. Not only do we achieve a de Sitter vacuum, but the scaling of ~\eqref{eq:Nscaling} corresponds to a large-volume scaling at the same time. since this scales down the size of the potential, see \eqref{eq:dSpotential}.

\subsection*{Higher order corrections with $G_4$-factors}
In the above construction we have been using $\mathcal{R}^{(4)}$ corrections together with a single $G_4$ component; $G_4$ is external since we have been using a flux that has its dual $G_7$ filling the internal space. The simultaneous use of these two terms may be problematic as the complete 8-derivative correction to the eleven dimensional supergravity is not known. This means that there are other terms that may become relevant. In this section we will outline an argument as to why these terms will not contribute.

Higher derivative corrections must, as mentioned before, enter via the Kähler potential. There are two ways that this can happen. The first is where a four dimensional Ricci-scalar can be peeled off from the correction term effectively correcting the four dimensional Planck-mass. By this we mean a correction term that has the form
\begin{equation}
\mathcal{R}^{(4)} = \mathcal{R}_4 \mathcal{R}^{(3)}\,,
\end{equation}
that corrects the Einstein-Hilbert term. The second is where a kinetic term can be peeled off from the correction term, and would lead to a contribution to the Kähler potential via the Kähler metric that defines the kinetic terms in the effective theory. Because of the underlying supersymmetry these corrections must come in a pair where both can recombine into a contribution to the Kähler potential. This means that if we find that there is no way of peeling off an $\mathcal{R}_4$ factor from any term in the full eight derivative correction there is no contribution from these terms to the Kähler potential at all.

Using this logic we can consider the explicit expressions that so far are known for eight derivative terms involving $G_4^{(2)}$ factors using that our $G_4$ only have external indices. If there is no term from which a Ricci scalar can be peeled off, we know that there is no contribution. For this purpose we have included the known terms in Appendix \ref{app:appendix}. There are four forms of contractions given by $t_8 t_8$, $\epsilon_{11}\epsilon_{11}$, $s_{18}$, and $Z$, that we will now consider in order.
\begin{itemize}
    \item[$t_8t_8$] For the $\mathcal{R}^{(4)}$ term, a Ricci scalar cannot be peeled off from the $t_8 t_8$ term, because any contraction of external indices are between different Riemann tensors. The same happens for the $t_8 t_8 G_4^{(2)}\mathcal{R}^{(3)}$, with the difference that Riemann tensors with external indices can also be contracted with $G_4$. Still, no Ricci scalar can be extracted.
    \item[$\epsilon_{11}\epsilon_{11}$] The $\epsilon_{11}\epsilon_{11} \mathcal{R}^{(4)}$ term does give rise to a Ricci scalar factor. The same cannot happen for $\epsilon_{11}\epsilon_{11} G_4^{(2)}\mathcal{R}^{(3)}$ however, since the external indices are reserved for $G_4$ as we only have an external component. Hence there are no remaining external indices to produce a Ricci scalar.
    \item[$s_{18}$] The $s_{18}$ is a tad bit more complicated as its full form has not been fully determined. However, considering Equation (B.19) of \cite{Weissenbacher:2019mef} which is a decomposition of terms that are part of the proposed expression for $s_{18}$, and we find no term where all indices of a single Riemann have been contracted on itself.
    \item[$Z$] Similarly in $Z$ there are only indices that are contracted between several Riemann tensors and hence a Ricci scalar is also absent here.
\end{itemize}
From this we conclude that our construction has no contributions from terms with a single $G_4^{(2)}$ factor. This is due to the fact that our $G_4$ fills spacetime. In any other configuration the $\epsilon_{11}\epsilon_{11} G_4^{(2)}\mathcal{R}^{(3)}$ term would have an $\mathcal{R}_4$ that could be extracted leading to a contribution to the Kähler potential consistent with contributions to the kinetic terms from all other eight derivative terms.

The above is only an analysis of the $G_4^{(2)}$, which is as explicit as we can be with the current knowledge of the form of the eight derivative corrections. In principle there are higher order terms in $G_4$ that could possibly contribute. We do find this unlikely, however, since bringing in more factors of $G_4$ implies that there are more external indices to be contracted, which was precisely why the $\epsilon_{11}\epsilon_{11}$ term was protected. Another factor of $G_4^{(2)}$ would bring in an additional eight external indices that has to be contracted in some way that, which for many terms would be Riemann tensors, again making it impossible to extract a Ricci scalar.

In the event that there would be a contribution found in the eight derivative terms with four $G_4$ or higher, or that one have been missed due to the possible incompletion of the $G_4^{(2)}$ contributions, we can still estimate the size these corrections would have. For the $\epsilon_{11}\epsilon_{11} \mathcal{R}^{(4)}$ term we simply recall
\begin{equation}
\sqrt{-g_{11}} \mathcal{R}^{(4)} \sim \sqrt{-g_{11}} \mathcal{R}_E^{(1)} \mathcal{R}_I^{(3)} \sim \rho^{-7/2} \rho^{7/2} \rho^{-3} \sim \rho^{-3} \sim \frac{1}{N^{2} \rho_0^{3}}\ ,
\end{equation}
where the $E/I$ labels refer to external and internal, respectively. The analogous scaling for terms with $G_4$ factors, where a Ricci scalar can be extracted, would be
\begin{equation}
\sqrt{-g_{11}} G_4^{(2m)}\mathcal{R}^{(4-m)} \sim \sqrt{-g_{11}} G_4^{(2m)} \mathcal{R}_E^{(1)} \mathcal{R}_I^{(3-m)} \sim \frac{1}{N^2 \rho_0^{3(1+2m)}}\ ,
\end{equation}
where $G_4$ it self scales as
\begin{equation}
G_4 \sim N^2 \rho^{-21/2}\ .
\end{equation}
This means that these terms would be as important as the $\mathcal{R}^{(4)}$ term when it comes to $N$, but have a different scaling with respect to $\rho$. If such terms are there, contrary to our expectation, our construction of a dS vacuum would need to be modified.
	% !TEX root = main.tex

\section{Scale hierarchies and loop corrections}
\label{sec:loop}

In this work we have provided a mechanism for constructing metastable dS vacua within M-theory by making use of higher derivative corrections.
Note that this means that our construction maps into a strongly coupled type IIA string background, and nevertheless, at least as far the higher derivative
expansion is concerned, we have shown how this is under control in our setup.
However, once we have obtained such a dS vacuum, we also need to discuss the relevance of quantum corrections and assess whether or not they affect the stability of the previously found vacuum.
A massive scalar with mass $m$ gives a contribution
\bea
\rho&=&\frac{1}{2} \int \frac{d^4 k_E}{(2\pi )^4}\log \left( \frac{k_E^2+m^2}{\mu^2} \right) \\
&=& \frac{1}{64 \pi^2} \left( \Lambda^4 \log \left(\frac{\Lambda^2}{\mu^2}\right) + 2\Lambda^2 m^2+  m^4 \log \left(\frac{m^2}{\Lambda^2} \right) \right) +...\ ,
\eea
to the vacuum energy if $m\ll\Lambda$, where the cutoff $\Lambda$ is covariant. See \eg \cite{Danielsson:2018qpa} for a review. In a theory with $\mathcal{N}=1$ supersymmetry summing over all fermions and bosons, the corrections are given by
the Coleman-Weinberg potential \cite{Coleman:1973jx}
\be
\delta V_{1\textrm{-loop}} = \frac{1}{64 \pi^2} \left[ \Lambda^4 \,\textrm{STr}(m^0) \log \left(\frac{\Lambda^2}{\mu^2}\right) + 2\Lambda^2 \,\textrm{STr}(m^2)+ \textrm{STr} \left(m^4\log \left(\frac{m^2}{\Lambda^2}\right) \right) \right]\ , \label{CW}
\ee
where the supertrace sums over all bosonic as well as all fermionic degrees of freedom with a weight $(-1)^{2j} (2j+1)$, where $j$ is the spin.
Even when supersymmetry is broken spontaneously at a scale set by the gravitino mass $m_{3/2}$, one still expects an improved UV behavior due to the structure of the corrections
dictated by supersymmetry.
In particular, the first term in the above expansion always vanishes due to the matching between the number of fermionic and bosonic dynamical degrees of freedom, while the second term, if present, behaves as $m_{3/2}^2 \Lambda^2$, and the third term as $m_{3/2}^4$.
As for the second term, it may or may not vanish depending on the actual details of a given phenomenological model. This will single out scenarios which are
physically more appealing based on naturalness arguments.
The appropriate choice for $\Lambda$ is the Kaluza-Klein mass, $m_{\textrm{KK}}$,
at least if the goal is that of retaining a reliable 4D effective description of our dS model.
In \cite{Cicoli:2007xp} this structure was reproduced in the case of various string theory compactifications on Calabi-Yau manifolds.

In our case there are some important differences.  We find that the cutoff is given by
\be
\Lambda \sim \frac{1}{\rho^{1/2}}\frac{1}{\rho^{7/4}} \sim \frac{1}{\rho^{9/4}} \sim N^{-3/2} \ ,
\ee
where the first factor accounts for the inverse KK radius in string frame, while the second factor is due to the change to 4D Einstein frame.
On the other hand, the gravitino mass is given by
\be
m_{3/2} \sim \frac{N}{\rho^{9/4}} \sim N^{-1/2} \ .
\ee
with $m_{3/2} \gg \Lambda$. Hence, supersymmetry is not restored before you reach the scale of the extra dimensions. Physically, this implies that even though the low energy theory is described by $\mathcal{N}=1$ supergravity, supersymmetry can never be restored in 4D. In particular, there are no 4D superpartners running in the loops and contributing to the renormalization of 4D physical observables. If we then repeat the calculation of the one loop
corrections bearing this in mind, we find the leading contribution to be simply given by $\Lambda ^4 \sim 1/\rho^9 \sim 1/N^6$, if we integrate out all the particles with masses $m$ less than $\Lambda=m_{\textrm{KK}}$, supersymmetry is no longer forcing any cancellations to take place. Note that \eqref{CW} is calculated assuming $m, m_{3/2} \gg \Lambda$, which no longer is true. We note that these corrections are much smaller than the classical terms, which are of order $1/N^3$. Hence, they cannot affect stability in our scenario where $N$ is taken to be large, even though they might possibly be of phenomenological importance.

To conclude, let us discuss the amount of fine-tuning needed in our model for producing values of the cosmological constant consistent with the current experimental observations.
In any realistic theory, we expect $ \Lambda^2 M_{\textrm{Pl}}^2\gg \Lambda^4 \gg \Lambda_{\textrm{cc}}$. This is a stronger requirement than scale separation, \ie $\Lambda\gg H$,
which would just imply $\Lambda^2 M_{\textrm{Pl}}^2\gg \Lambda_{\textrm{cc}}$. Therefore, we need to fine-tune the contribution to the vacuum energy produced by the classical theory so that you obtain the observed value of the cosmological constant once the loop corrections are added. Depending on the sign of the loop correction, the classical vacuum could turn out to be AdS or dS. In our numerical example we have $V=0.006/N^3$, where the small prefactor guarantees scale separation. To make our model fully realistic we need, at fixed $N$, arrange for an accidental fine-tuning making the classical contribution so small that it can be almost canceled by the quantum contribution of size $1/N^6$, leaving the observed value of the cosmological constant. We have checked this numerically and there appear to be no obstruction against such fine-tuning in our model.

A possible caveat that has recently been discussed though we have not considered here, is the choice of vacuum when performing the loop calculation. The results in \cite{Berg:2005yu,Berg:2007wt,Cicoli:2007xp}, using Euclidean continuations, implicitly assume the Bunch-Davies vacuum \cite{Bunch:1978yq}. More precisely, they assume that the quantum state is
such that it approaches the Minkowski vacuum at high energies. As discussed in \eg \cite{Mottola:1984ar,Danielsson:2002kx,Polyakov:2007mm,Polyakov:2009nq,Krotov:2010ma,Polyakov:2012uc,Anderson:2013ila,Anderson:2013zia,Anderson:2017hts}, and recently in \cite{Danielsson:2018qpa} with further references, it is far from clear that this has to be the case, or even can be the case. Particularly in the physically realistic regime, where $\Lambda^4 \gg \Lambda_{\textrm{cc}}$, a possible time evolution of the dark energy will crucially depend on the physics of the quantum corrections and how they might adjust and shift the value of the dark energy, possibly in a time dependent way. This is true even if they remain small and cannot effect stability. These issues need to be understood if one wants to judge our proposal, or any other proposal, of a dS vacuum.
Should these issues turn out to have dramatic consequences for any description of a dS vacuum, then one would have to rethink of completely new ways of conceiving dark energy and cosmology. A possibly promising alternative would be to holographically resolve time dependence in one dimension higher, as it has been recently discussed in \cite{Banerjee:2018qey,Heckman:2018mxl,Heckman:2019dsj}.

    % !TEX root = main.tex

\section*{Acknowledgments}
We would like to thank Fridrik Freyr Gautason and Raffaele Savelli for their input.
The work of UD, GD and SG is supported by the Swedish Research Council (VR) and the work of JB is supported by MIUR-PRIN contract 2015MP2CX4002 ``\emph{Non-perturbative aspects of gauge theories and strings}''.

%	\appendix
    \begin{appendices}
	% !TEX root = main.tex

\section{Eight derivative corrections}
\label{app:appendix}

The eleven dimensional supergravity action receives corrections from higher derivative terms, starting at eight derivatives, as mentioned previously. In this appendix we review the known explicit form of a subset of these contributions which
are closed under supersymmetry transformations.

To the degree that these terms are known, there are contributions of the form
\begin{equation}
\delta  S \sim \int_{11} \sum_{m=0}^{4} \mathcal{R}^{(4-m)} G_4^{(2m)} + \sum_{m=1}^{4} \mathcal{R}^{(3-m)} (\nabla G_4)^{(2m)} + C_3 \w X_8\ ,
\end{equation}
where the power indicate some contraction of the various indices. The explicitly know terms have the contraction composed as \cite{Hyakutake:2006aq,Liu:2013dna}
\begin{equation}
  \delta S|_{\mathcal{R}^{(4)}} = \frac{1}{2 \kappa_{11}^2} \int \star_{11} \left(t_8t_8 - \frac{1}{24} \epsilon_{11}\epsilon_{11}\right)\mathcal{R}^{(4)} + C_3 \w X_8\ ,
\end{equation}
where the index contractions symbolised here by $t_8$ and $\epsilon_{11}$ are
\begin{equation}
\label{eq:explicit_R4}
  \begin{split}
    t_8 t_8 \mathcal{R}^{(4)} &= t_8^{M_1 \ldots M_8} {t_8}_{N_1 \ldots N_8} \mathcal{R}^{N_1 N_2}_{M_1 M_2} \ldots \mathcal{R}^{N_7 N_8}_{M_7 M_8}\ ,\\
    \epsilon_{11} \epsilon_{11} \mathcal{R}^{(4)} &= \epsilon^{A_1 A_2 A_3 M_1 \ldots M_8} \epsilon_{A_1 A_2 A_3 N_1 \ldots N_8} \mathcal{R}^{N_1 N_2}_{M_1 M_2} \ldots \mathcal{R}^{N_7 N_8}_{M_7 M_8}\ ,\\
    X_8 &= \frac{1}{(2\pi)^4} \frac{1}{192} \left(\Tr {R}^4 - \frac{1}{4} (\Tr {R}^2)^2\right)\ ,
  \end{split}
\end{equation}
where $R$ is the curvature two-form $R^M_{\ P} = \mathcal{R}^M_{\ PQR}\, \d x^Q \w \d x^R$. While $\epsilon_{11}$ is the eleven dimensional Levi-Civita tensor, the $t_8$ acts on antisymmetric tensors as
\begin{equation}
  t_8^{M_1 \ldots M_8} A_{M_1 M_2} \ldots A_{M_7 M_8} = 24 \Tr A^4 - 6 (\Tr A^2)^2\,.
\end{equation}

For the terms with $G_4^{(2)}$ factors, the expressions are less known, see e.g.~\cite{Weissenbacher:2019mef}, where a full expression for the $G_4^{(2)}$ terms are proposed for Calabi-Yau four-folds. Even though a fully general form for these is not yet known, especially not on the geometry we have considered, we will not need any of these details in this work. The proposed expression is
\begin{equation}
  S|_{{G}_4^{(2)}} = \frac{1}{2\kappa^2_{11}} = \int_{11} - \star \left(t_8 t_8 + \frac{1}{96} \epsilon_{11} \epsilon_{11}\right) G_4^{(2)} \mathcal{R}^{(3)} + \star s_{18} (\nabla G_4)^{(2)} \mathcal{R}^{(2)} + 256 Z G_4 \wedge \star G_4\ ,
\end{equation}
where $s_{18}$ is some complicated contraction involving Riemann tensors and $G_{4}$'s which may be found in Appendix B of \cite{Weissenbacher:2019mef}. The $Z$ is given by \cite{Grimm:2017okk}
\begin{equation}
  Z = \frac{1}{12} \left( \mathcal{R}_{M_1 M_2}^{\qquad M_3 M_4} \mathcal{R}_{M_3 M_4}^{\qquad M_5 M_6} \mathcal{R}_{M_5 M_6}^{\qquad M_1 M_2} - 2 \mathcal{R}_{M_1\ \ M_3}^{\ \ M_2\ \ M^4} \mathcal{R}_{M_2\ \ M_4}^{\ \ M_5\ \ M^6} \mathcal{R}_{M_5\ \ M_6}^{\ \ M_1\ \ M^2} \right)\ .
\end{equation}
The $t_8$ and $\epsilon_{11}$ contractions are here
\begin{equation}
  \begin{split}
    t_8 t_8 G_4^{(2)} \mathcal{R}^{(3)} &= t_8^{M_1 \ldots M_8} {t_8}_{N_1 \ldots N_8} G^{N_1}_{\ \ M_1 A_1 A_2} G^{N_2\ \ A_1 A_2}_{\ \ M_2} \mathcal{R}^{N_3 N_4}_{\qquad M_3 M_4} \mathcal{R}^{N_5 N_6}_{\qquad M_5 M_6} \mathcal{R}^{N_7 N_8}_{\qquad M_7 M_8}\ ,\\
    \epsilon_{11} \epsilon_{11} G_4^{(2)} \mathcal{R}^{(3)} &= \epsilon^{A M_1 \ldots M_{10}} \epsilon_{A N_1 \ldots N_{10}} G^{N_1 N_2}_{\qquad M_1 M_2} G^{N_3 N_4}_{\qquad M_3 M_4} \mathcal{R}^{N_5 N_6}_{\qquad M_5 M_6} \mathcal{R}^{N_7 N_8}_{\qquad M_7 M_8} \mathcal{R}^{N_9 N_{10}}_{\qquad M_9 M_{10}}\ .
  \end{split}
\end{equation}
The higher $G_4^{(m)}$ expressions are even less known, and we are not aware of any established contractions for these.
Nevertheless, we argued in Section~\ref{sec:ds} that the form of these terms is not crucial to our analysis.

    \end{appendices}

  % References
  \small
  \bibliography{references}
  \bibliographystyle{utphysmodb}
\end{document}